\documentclass[aps,pra,10pt,twocolumn,superscriptaddress]{revtex4}
\usepackage{amsmath}
\usepackage{amssymb}
\usepackage{bbm}
\usepackage{graphicx}
\usepackage{framed}
\usepackage{color}
\usepackage{hyperref}
\usepackage{epstopdf}
\usepackage{dsfont}
\usepackage{amssymb}
\usepackage{amsmath}
\usepackage{amsthm}
\usepackage{wrapfig}
\usepackage{relsize}
\usepackage{bm}
\usepackage[latin1]{inputenc}
\usepackage{enumitem}

\newcommand{\Tr}{{\rm Tr}}
\newcommand{\red}[1]{{\color{red}}}

\def\Tr{\hbox{Tr}} 
\usepackage{pifont}

\newcommand{\ket}[1]{\vert#1\rangle}
\newcommand{\bra}[1]{\langle#1\vert}

\newcommand{\ba}{\begin{eqnarray}}
\newcommand{\ea}{\end{eqnarray}}

\begin{document}

\title{Speed of qubit states during thermalisation} 

\author{Michele M. Feyles}
\affiliation{Dipartimento di Fisica, Sapienza Universit\`{a} di Roma, P.le Aldo Moro 5, 00185, Rome, Italy}

\author{Luca Mancino}
\affiliation{Centre for Theoretical Atomic, Molecular and Optical Physics, School of Mathematics and Physics, Queen's University Belfast, Belfast BT7 1NN, United Kingdom}
\affiliation{Dipartimento di Scienze, Universit\`{a} degli Studi Roma Tre, Via della Vasca Navale 84, 00146, Rome, Italy}

\author{Marco Sbroscia}
\affiliation{Dipartimento di Scienze, Universit\`{a} degli Studi Roma Tre, Via della Vasca Navale 84, 00146, Rome, Italy}

\author{Ilaria Gianani}
\affiliation{Dipartimento di Scienze, Universit\`{a} degli Studi Roma Tre, Via della Vasca Navale 84, 00146, Rome, Italy}

\author{Marco Barbieri}
\affiliation{Dipartimento di Scienze, Universit\`{a} degli Studi Roma Tre, Via della Vasca Navale 84, 00146, Rome, Italy}
\affiliation{Istituto Nazionale di Ottica - CNR, Largo Enrico Fermi 6, 50125, Florence, Italy}

\begin{abstract} 
Classifying quantum states usually demands to observe properties such as the amount of correlation at one point in time. Further insight may be gained by inspecting the dynamics in a given evolution scheme. Here we attempt such a classification looking at single-qubit and two-qubit states at the start of thermalisation with a heat bath. The speed with which the evolution starts is influenced by quantum aspects of the state, however, such signatures do not allow for a systematic classification.
\end{abstract}\maketitle

{\it Introduction.} Correlations among systems provide the most spectacular departure of quantum mechanics from our ordinary experience. The comparison of the outcomes of local observables might reveal connections defying notions of classicality. In the case of bipartite systems, a systematic classification can be carried out, identifying what exact aspect is considered: discord~\cite{Ollivier01}, entanglement~\cite{Vedral97,Horodecki09}, steerability~\cite{Wiseman07,Milne14}, nonlocality~\cite{Bancal09,Brunner14}, or work extraction~\cite{Maruyama05,Viguie05,Ciampini17}.  All these provide a characterisation of the quantum state at a given time, regardless the evolution it is undergoing. Entanglement has been recognised as a resource to achieve quantum speed limits unattainable with separable states \cite{Giovannetti02}; also, it has been demonstrated how entanglement may be consumed in a finite time even in quantum channels for which the cancellation of single-system properties is asymptotic \cite{Yu04,Yu06,Santos06}. These seem to suggest that entanglement might be manifested also in \textit{dynamical} properties, \textit{i.e.} those connected to the evolution of the state. In this respect, there has been a recent interest in studying such evolutions on the Riemannian manifold of density matrices~\cite{Braunstein94}. The interest in these methods partly originated from the explicit link between the metric on the manifold and informational quantities such as the Fisher information~\cite{Pires16}. Applications of geometry has lead to investigating relations between quantum quirks and Riemannian metrics \cite{Pires15,Bej18}, bounds to the entropy production in closed and open systems~\cite{Deffner10,Mancino18}, and establishing more general quantum speed limits than those previously known~\cite{Pires16,delCampo13}.  

A dynamical quantity of particular interest is the speed of the states on this manifold \cite{Anandan1990,Taddei2013,Cianciaruso2017}. In the case of unitary transformations, it is possible to use such speed to derive a useful criterion to witness entanglement~\cite{Pezze16}. In this work, we inspect how the speed of a quantum state over the manifold is affected by coherence and correlation properties during dissipative dynamics: the case in point is the thermalisation of single-qubit and two-qubit systems in contact with a bosonic thermal bath at fixed temperature. Our results show that the link is subtle and quantitative bounds are elusive, but some general considerations can nevertheless be drawn. 

{\it Geometric considerations.} Complying with what is commonly established in quantum mechanics, a quantum system is defined in a Hilbert space $\mathcal{H}$, and the set of its states forms a Riemannian manifold $\mathcal{S}=\mathcal{D}(\mathcal{H})$ of density matrices over $\mathcal{H}$. Riemannian metrics can be defined associating an infinitesimal length $ds^2=\Gamma_\rho(d\rho,d\rho)$ between the density operators $\rho$ and $\rho+d\rho$. The choice of $\Gamma_\rho$ is not unique: according to the Morozova, \v Cencov, and Petz (MCP) theorem \cite{Morozova91,Petz96,Petz296,Petz02,Hiai09}, each Riemannian metric is characterizable through a one-to-one correspondence with the set of the Morozova-\v Cencov (MC) functions $f(t):\mathbb{R}_{+} \rightarrow \mathbb{R}_{+}$. As shown in \cite{Kubo80}, a MC function has to satisfy $f_{m}(t) \leq f(t) \leq f_{M}(t)$, in which $f_{m}(t)=2t/(1+t)$, and $f_{M}(t)=(1+t)/2$ which leads to the Bures-Uhlmann metric, also known as quantum Fisher information metric \cite{Uhlmann93,Uhlmann95}. When considering a system undergoing an evolution that smoothly changes a set of parameters $\lbrace \lambda \rbrace_{i}$ characterising the quantum state, the infinitesimal length element can be written as $ds^2=\sum_{\mu \nu}G_{\mu \nu}d\lambda_\mu d\lambda_\nu$, where $d\lambda_\mu$ is the variation of the $\mu$-th element of the parameter set. For any given state $\rho=\sum_j p_j \vert j \rangle \langle j \vert$, there exists a decomposition:
\begin{equation}
G_{\mu \nu}^f = F_{\mu \nu}+Q_{\mu \nu}^f,
\label{metrica}
\end{equation}
where $F_{\mu \nu}=1/4 \sum_j \left( \partial_\mu p_j \partial_\nu p_j \right) /p_j$, and $Q^f_{\mu \nu}= - 1/2 \sum_{j<l} c^{f}(p_j,p_l)(p_j-p_l)^2 \bra{j}\partial_\mu\ket{l} \bra{l}\partial_\nu\ket{j}$. Here, $c^f (u,w)$ is a symmetric function $c^f (u,w)=c^f (w,u)$, derived from the so-called MC function $f(t)$, which satisfies the properties i) $c^f (\kappa u, \kappa w)=\kappa^{-1} c^f(u,w)$, and ii) $c^f (u,w)=1/(w f(u/w))$, so that is does not assume a specific role in defining the term $F_{\mu \nu}$ of Eq \eqref{metrica}. 

\begin{figure*}[t!]
\includegraphics[width=1\textwidth]{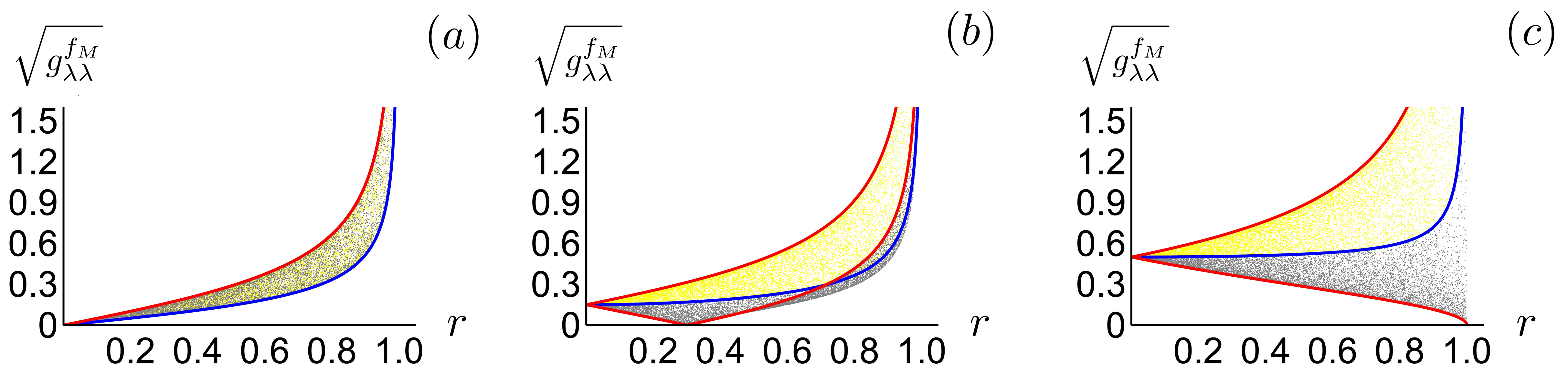}
\caption{Behaviour of the $\sqrt{g_{\lambda \lambda}^{f_M}}$ quantity for single-qubit states associated to different temperatures (Panel (a): $\alpha=0$, Panel (b): $\alpha=0.3$, Panel (c): $\alpha=1$) {\it vs} the purity of the state $r$. Blue curve: ``classical" states $x(0)=0$; red curve: states with $x(0)=1$. The points are generated from an even random distribution according to the Haar metric. States are differentiated according to their energies: $ Z>0$ (yellow points) and $\ Z \leq0$ (grey points).}
\label{Figure1}
\end{figure*}

{\it Single-qubit evolution.} The dynamics  considered here refers to the interaction of qubits with an external bosonic thermal bath: thermalisation occurs with a typical time scale $\eta$ that depends on the temperature \cite{Breuer02}. For the single qubit, we use the following parametrisation for the density matrix in terms of $\lambda = 1-e^{-\eta t}$: $\rho(\lambda)=1/2 \left( I_2 + z(\lambda) \sigma_z+ x(\lambda) \sigma_x \right)$, where $z(\lambda)$ is related to the way the energetic levels are populated, while $x(\lambda)$ accounts for contributions of quantum coherence \cite{Nielsen2000}. Here, the terms associated to $\sigma_y$ are not taken into account since the evolution of the system does not depend on the initial phase of the state being considered. Once the system approaches the fixed point of its dynamics, its state can be parametrised as $\rho_\infty=\rho(1)=1/2 \left( I_2 + \alpha \sigma_z \right)$: since $\lambda$ is the only parameter which changes during thermalisation, it is possible to fix $\alpha=z(1)$ as the population unbalance of the Gibbsian state at the temperature of the external reservoir. Here, we use $\psi(\lambda)=\rho(\lambda)-\rho_\infty=1/2 \left( Z(\lambda)\sigma_z + X(\lambda)\sigma_x \right)$ as a way to quantify how much the evolved density matrix of the system is \textit{different} from its thermalised one: in this scheme $Z(\lambda)=z(\lambda)-\alpha=Z(0)(1-\lambda)$ and $X(\lambda)=X(0)\sqrt{1-\lambda}$, where $Z(0)=\left( z(0) - \alpha \right)$, and $X(\lambda)=x(\lambda)$ for $\lambda \in \left[ 0, 1 \right]$, reminiscent of the different longitudinal and transverse decay times. 


With the aim to establish the way the quantum properties of the considered state vary during the system dynamics, we make use of Riemannian geometry over $\mathcal{S}$. From a geometric point of view, the evolution of the state associated to the thermalisation dynamics of the system draws a path $\gamma$ onto the Riemannian manifold of the density matrices with length $l_\gamma^f(\tau)=\int_0^\tau dt \sqrt{\sum_{\mu \nu} G_{\mu \nu}^f (\partial_t \lambda_\mu) (\partial_t \lambda_\nu)}$ \cite{Pires16}; here, the integrand refers to the Riemannian speed $V(\rho)$ over $\mathcal{S}$. In our case, such a quantity reduces to
\begin{equation}
V(\rho)=\sqrt{G_{tt}^f},
\end{equation}
which can be also expressed as $V(\rho)=\sqrt{G_{\lambda \lambda}^f \eta^2} (1-\lambda)$. According to this matching, it is possible to introduce the following relation
\begin{equation}
G_{\lambda \lambda}^f = g_{\lambda \lambda}^f (1-\lambda)^{-2}
\label{rep}
\end{equation}
where $g_{\lambda \lambda}^f=G_{tt}^f/\eta^2$ does not show an explicit dependence on time, and can be also decomposed in the two contributions $\varphi_{\lambda \lambda}$ and $q^f_{\lambda \lambda}$ following Eq.\eqref{metrica} where the two terms bear analogous meaning. More in detail, the expressions read as
\begin{equation}
\begin{aligned}
\varphi_{\lambda \lambda} &= \frac{1}{4} \frac{(z\,\partial_\lambda z + x\, \partial_\lambda x)}{r^2(1-r^2)},\\
q^f_{\lambda \lambda} &= \frac{1}{8} c^{f}(p_1,p_2)\frac{(x\, \partial_\lambda z - z \, \partial_\lambda x)}{r^2},
\end{aligned}
\label{metrica2}
\end{equation}
where $r^2 = x^2+z^2$, and $p_1, p_2$ are the eigenvalues of $\rho$. For the sake of clarity, the dependence on $\lambda$ is omitted. 

A way to characterize $g_{\lambda \lambda}^f$ or explicitly its $q_{\lambda \lambda}^f$ part in Eq.\eqref{metrica2} consists in defining a metric onto the Riemannian manifold. In what follows, we adopt the choice made in \cite{Pezze16} where Fisher metric has been used: as established by the MCP theorem, this is tantamount to choose $f_M(t)$ as the considered MC function, thus leading to $c^{f_M}(p_1,p_2)=2$. As a consequence of this choice, the length separating two states $\rho_1$ and $\rho_2$ is directly linked with their fidelity ${\mathcal F}(\rho_1,\rho_2) = (\Tr[\sqrt{\sqrt{\rho_1}\rho_2\sqrt{\rho_1}}])^2$ \cite{Pires16}, and
\begin{equation}
g^{f_M}_{\lambda \lambda} = \frac{1}{4} \frac{Z^2 + X^2/4(1-(Z-\alpha)^2)}{(1-r^2)}.
\end{equation}
We adopt the initial speed (i.e. that at $\lambda=0$) as our figure to assess dynamical properties. We notice that $g^{f_M}_{\lambda \lambda}$ diverges at $\lambda=0$ for all pure states, independently on the presence of coherence. The behaviour of this quantity (that at $\lambda=0$) is reported in Fig.\ref{Figure1} as a function of the purity $r$ of the initial state $\rho(0)$ for different temperatures. Notice that the speed diverges for pure states, regardless the initial presence of coherence.


\begin{figure*}[t!]
\includegraphics[width=1\textwidth]{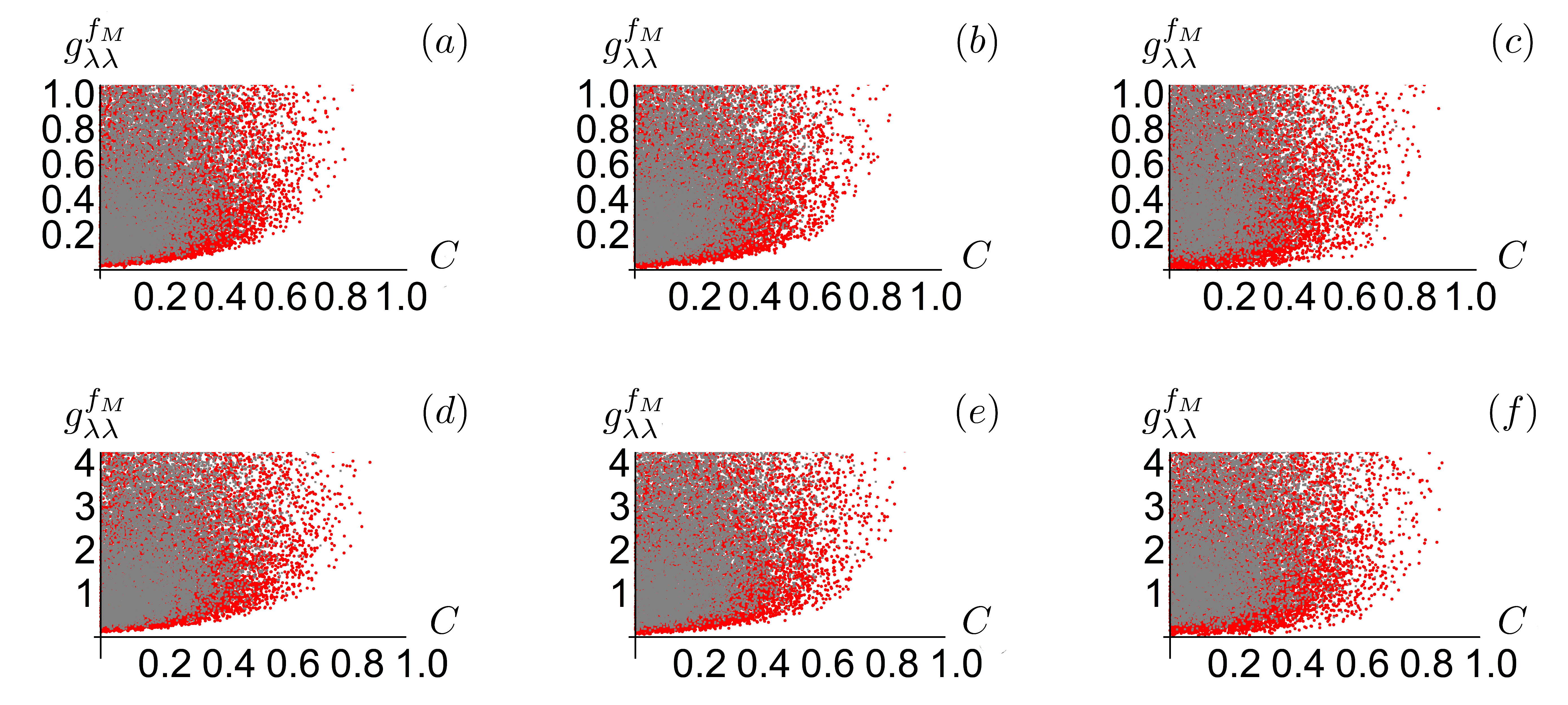}
\caption{Behaviour of the $g_{\lambda \lambda}^{f_M}$ quantity for two-qubit states systems with thermalising maps acting on either a single (a-c) or both (d-f) subsystems, {\it vs} the concurrence $C$ of the same states. The panels refer to different thermalisation temperatures: in detail, $\alpha=0$ for Panels (a) and (d), $\alpha=0.5$ for Panels (b) and (e), and $\alpha=1$ for Panels (c) and (f). Red points stand for $X$ states, while gray points stand for generic bipartite states; all the points have been generated from an even random distribution according to the Haar metric.}
\label{Figure2}
\end{figure*}

At infinite temperature, the fixed point is at the centre of the Bloch sphere, hence thermalisation seconds its symmetry: at a given distance from $\rho_\infty$, the initial speed $g^{f_M}_{\lambda \lambda}(t=0)$ shows a neat dependence on the angle on the Bloch sphere. The states with $z(0)=0$, whose dynamics is entirely dictated by the relaxation of coherence start slower than those for $x(0)=0$, whose dynamics only presents population relaxation.  At finite bath temperatures,  the asymmetry of the fixed point determines a non trivial interplay between these two factors: while the fastest states at given distance are still those with $z(0)\geq \alpha$, the form of the slowest states depends on the value of $r$.  As a general remark, the closer the state to the fixed point $\rho_\infty$, the slower it will start moving; moreover, a series development at long times $\lambda\sim1$ reads $g^f_{\lambda \lambda}\sim X(0)^2(1-\lambda)/16+O((1-\lambda)^2)$, independent of $\alpha$, although this is a peculiarity of the chosen Fisher metric \cite{Appendix}.

So far we have focussed our attention on statistical speeds in the interaction picture: this is equivalent to neglecting the rotational contribution due to the free Hamiltonian $H={1 \over 2} \omega \sigma_z$ of the qubit. If we include the effects of the Hamiltonian on the speed, we find that for the single qubit the relation between Schr\"odinger and Interaction picture is rather simple and reads:
  \begin{equation}
  \label{orto}
  V^2_{S}(\rho)=V^2_{I}(\rho)+V^2_{H}(\rho)
  \end{equation}
where $V_{H}$ stand for the statistical speed of a state $\rho$ evolving with Hamiltonian $H$, and $V_{I}$ ,$V_{S}$ are the statistical speeds in Interaction and Schr\"odinger picture respectively. This is a consequence of the form of the metric, which as the same orthogonality relations as the euclidean metric under rotations, and the form of the thermalisation map, which is symmetric under rotations around the $z$ axis.

{\it Considerations on two-qubit systems.} The analysis above permits to identify the behaviour of the statistical speed of single-qubit systems interacting with an external bosonic thermal bath. When considering composite systems, the general form of the metrics and the statistical speed over the Riemannian manifold $\mathcal{S}$ become extremely complicated: these two end up depending on too many parameters to get any insight directly from the formulae; furthermore, correlation quantifiers are usually nonlinear functions of the same parameters. For these reasons, we deemed it more appropriate to investigate even simple two-qubit systems with numerical methods. However, proving numerically what is true is extremely hard, while proving what is not is much simpler~\footnote{M. Feyles deserves full credit for this wonderful sentence.}. We have been able to provide negative - \textit{but insightful} - answers to three interesting and natural questions.

\begin{figure*}[t!]
\includegraphics[width=1\textwidth]{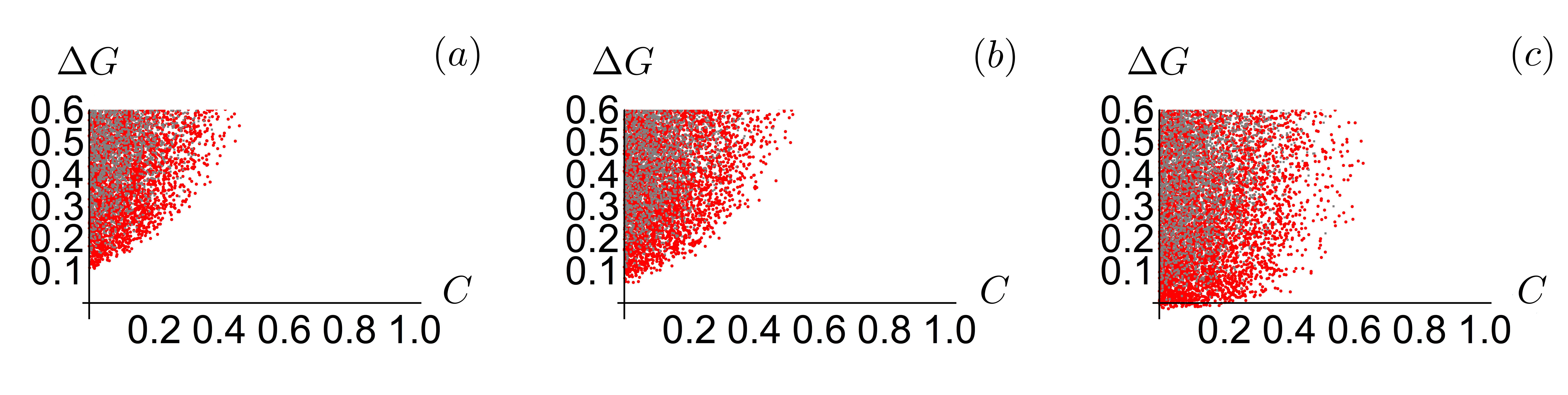}
\caption{Behaviour of $\Delta G=G_{\lambda \lambda}^{AB}-G_{\lambda \lambda}^{A}-G_{\lambda \lambda}^{B}$ for two-qubit systems with thermalising maps acting on both subsystems, {\it vs} the concurrence C of the same states. The panels refer to different thermalisation temperatures: in detail, $\alpha=0$ for Panel (a), $\alpha=0.5$ for Panel (b), and $\alpha=1$ for Panel (c) in which our numerical analysis has permitted to identify states with $\Delta G \leq 0$. Red points stand for $X$ states, while gray points stand for generic bipartite states; all the points have been generated from an even random distribution according to the Haar metric.}
\label{Figure3}
\end{figure*}

The first aspect we address is the effect of entanglement on the statistical speed: based on the results mentioned in the introduction, one may expect qualitatively different behaviours for entangled states with respect to separable states. The plots in Fig.~\ref{Figure2} tell us otherwise: these report the initial speed {\it vs} the concurrence $C$ of the states, for thermalising maps acting on either a single subsystem or both. In the latter case, we considered two independent maps at the same temperature for the sake of simplicity. One can see that, on average, entangled states reach higher speeds: this, however, seems to be linked to the purity of such states, as well as their distance from the final thermalised state, rather than entanglement {\it per se}. Direct signatures, if existing, should be searched elsewhere.

In such a context, we proceed with a simple observation: an almost pure entangled state will have its partitions in a mixed state, therefore their reduced evolutions will start with a term $G^{A}_{\lambda \lambda}, G^{B}_{\lambda \lambda}$ considerably smaller than those of the complete system $G^{AB}_{\lambda \lambda}$. We can then address the question as to whether a quantitative bound can be put on entangled states based on these considerations. Figure~\ref{Figure3} reports values of $\Delta G=G^{AB}_{\lambda \lambda}-G^{A}_{\lambda \lambda}-G^{B}_{\lambda \lambda}$, and reveals that such effort leads to nowhere: the presence of entanglement is not liked quantitatively to a discrepancy in global and local speeds. Negative values of $\Delta G$ can be obtained for states which are diagonal in the energy basis, but also for Werner entangled states: thus, this quantity cannot be used as a flag for classical correlations.

Finally, we remark that the simple property \eqref{orto} connecting the speeds in the Schr{\"{o}}dinger and interaction schemes does not hold for higher dimensions. Counter examples can be found by considering block matrices, covering different classes of states \cite{Appendix}. Inspecting their infinitesimal displacement, considering dissipation and a unitary Hamiltonian rotation at once, reveals how the orthogonality of the two contribution is not ensured as for the $Y_b$ class. Remarkably, orthogonality of the infinitesimal displacement, hence a relation similar to \eqref{orto} can be found for ``classical" states and specific entangled states: as in the qubit case, this is not specifically attached to a manifestation of quantum properties, nor lack thereof.

{\it Conclusions.} Inspecting the metrics of quantum states under unitary evolution can provide in many cases a neat categorisation of quantum states. The same efforts, carried out for dissipative dynamics, lead to results hard to interpret, given the many factors coming into play. The case of single qubits we have studied still provide a clear picture, indicating that quantum coherence does have a role in determining the initial thermalisation speed, however geometrical considerations, {\it viz.} the distance from the fixed point, seem to dominate. In the case of two qubits, a certain parallelism still holds, however, a prominent property such as quantum entanglement manifests in the speed only in a marginal way. Should we conclude that this is a manifestation of the fact thermalisation is a very classical problem, at a difference to unitary processes? This is likely to be so, as classical correlations become responsible for specific signatures in the speed. However, the lesson to be taken is that the problem has so many facets that isolating a particular one over the others can only lead to distorted and unsatisfactory pictures.

{\it Acknowledgements.} We are grateful to V. Cavina, A. De Pasquale, E. Roccia, V. Cimini, B. Capone, and M. Paternostro for useful discussions. LM acknowledges financial supports from the Angelo Della Riccia Foundation.

\subsection*{APPENDIX}

{\it Contractive metrics.} The $n \times n$ parameters characterizing the state of a system forms a Riemannian manifold $\mathcal{S}=\mathcal{D}(\mathcal{H})$ of density operators over the Hilbert space $\mathcal{H}$. Accordingly to what has been described in the main text, a spectrum of metrics can be defined on such a smooth manifold: a general definition of a metric is an application $\Gamma_\rho(*,*)$ that associates a real number to any zero-trace complex matrices $A$, $B$ $\in \mathcal{M}_n^0(\mathbb{C})$ \footnote{These conditions can be generalised, although at the expense of the compactness of the expressions. This restricted form is sufficient for our puroposes.} for any $\rho \in  \mathcal{M}_n$, the space of  $n\times n$ density matrices. These applications need to satisfy the following properties:
\begin{enumerate}
\item $(A, B) \to \Gamma_\rho( A, B)$ is bilinear;
\item $\Gamma_\rho(A, A) \geq 0$, with the equality holding only when $A$ is the null matrix;
\item $\rho \to \Gamma_\rho(A, A)$ is continuous on $\mathcal{M}_n$ for any $A$.
\end{enumerate}

A further request defines the property of monotonicity: 4. $\Gamma_{T(\rho)}(T( A), T( A)) \leq \Gamma_\rho(A, A)$ for any stochastic map $T$.

The MCP theorem \cite{Morozova91,Petz96,Petz296,Petz02,Hiai09} ensures that all contractive metrics can be put in the form:
 \begin{equation}
\Gamma^f_\rho(A,B)=\frac{1}{4} tr \left[A\, {c^f(\textbf{L}_\rho,\textbf{R}_\rho)}\,B \right],
\end{equation}
where $\textbf{L}_\rho,\textbf{R}_\rho:\mathcal{L}(\mathcal{H}) \rightarrow \mathcal{L}(\mathcal{H})$ are two linear super-operators which are defined on the space of the linear operators $\mathcal{L}(\mathcal{H})$ over $\mathcal{H}$. These satisfy the following relations: $\textbf{L}_\rho A = \rho A$, and $\textbf{R}_\rho A = A \rho$. The $c^f(u,w)$ function, which has been also described in the main text, depends on the MC function $f(t)$ and permits to define a specific metric onto the manifold $\mathcal{S}$. Specifically, we have adopted $f^M(t)=(1+t)/2$ which leads to the Bures-Uhlmann metric; within the whole spectrum of available $f(t)$, there are only two cases for which the explicit expressions for the geodesic distances are known: the first is represented by $f^M(t)$, and the second is associated to $f^{WY}(t)=1/4 (\sqrt{t}+1)^2$ leading to the Wigner-Yanase metric.

The infinitesimal displacement can be cast as  $ds^2=\Gamma^f_\rho(d\rho, d\rho)$, where $d\rho \in T_\rho [\mathcal{S}]$, the tangent space to that of the density matrices at $\rho$. The eigenbasis of $\rho$ $\left(\rho=\sum_i p_i \ket{i} \bra{i}\right)$ allows for the following expression:
\begin{equation}
\label{ds_contrattive}
ds^2=\frac{1}{4}\left[ \sum_i \frac{(d\rho_{ii})^2}{p_i} + 2 \sum_{j<i}c^f(p_i, p_j) \vert d\rho_{ij}\vert^2 \right].
\end{equation}
It is evident that for any ``classical'' state, the metric is reduced to its classical Fisher information. Given a parametrisation $\{\lambda_\mu\}$ of the state,
hence $d\rho=\sum_\mu \partial_\mu\rho\,d\lambda_\mu$, the following expression can be obtained
\begin{equation}
ds^2=\sum_{\mu \nu}G^f_{\mu \nu} d\lambda_\mu d\lambda_\nu.
\end{equation}
The $G_{\mu \nu}^f$ term can be decomposed in two different contributions $F_{\mu \nu}$ and $Q_{\mu \nu}^f$, as shown in Eq.\eqref{metrica}, which explicitly appear as 
\begin{equation}
\label{QF_F}
F_{\mu \nu}=\frac{1}{4} \sum_i \frac{\partial_\mu p_i  \partial_\nu p_i}{p_i},
\end{equation}
and
\begin{equation}
\label{QF_Q}
Q^f_{\mu \nu}=\frac{1}{2} \sum_{i<j} c^f(p_i,p_j)(p_i-p_j)^2 \mathcal {A}_{ji}^\mu \mathcal {A}_{ij}^\nu,
\end{equation}
where $\mathcal {A}_{jl}^\mu= i \bra{j}\partial_\mu \ket{l}$.

{\it Details on thermalisation.} Within the whole analysis, we have considered qubit systems - either isolated or in pairs - in contact with a bosonic thermal bath. The Hamiltonian of each qubit can be taken in the form $H=\frac{1}{2} \omega \sigma_z$, where $\ket{0}$ denotes the excited state, and $\ket{1}$ the ground state. The representation of a generic density matrix $\rho$ in terms of Pauli operators allows to use the length of the Bloch vector $r^2=x^2+y^2+z^2 \leq 1$ as a measure of the purity of the state;  the triplet $\{r,\theta,\phi\}$ offers an alternative choice for the coordinates of $\rho$, where $\theta$ and $\phi$ denote the polar and the azimuthal angle over the Bloch sphere. From a mathematical point of view, the interaction process which takes place when the system comes in contact with the bosonic thermal bath can be described by a two-parameters family of quantum channels $\lbrace \Phi_\lambda^p \rbrace$ such that $\rho(\lambda,p)=\Phi_\lambda^p [\rho_0]$, where $\rho_0$ stands for the initial state of the considered system. More specifically, the evolved state can be formally expressed in terms of a Generalized Amplitude Damping (GAD) channel, with Kraus operators \cite{Mancino18,Breuer02,Nielsen2000}: $E_0=\sqrt{p}(\vert 0 \rangle \langle 0 \vert + \sqrt{1-\lambda} \vert 1 \rangle \langle 1 \vert)$, $E_1=\sqrt{p} \sqrt{\lambda} \vert 0 \rangle \langle 1 \vert$, $E_2=\sqrt{1-p}(\sqrt{1-\lambda} \vert 0 \rangle \langle 0 \vert + \vert 1 \rangle \langle 1 \vert)$, $E_3=\sqrt{1-p}\sqrt{\lambda} \vert 1 \rangle \langle 0 \vert$. Here $p$ and $\lambda=1-e^{-\eta t}$ represent respectively the temperature- and time-dependent probability and damping coefficients. The evolution under the Kraus operators leads to the following transformation: 
\begin{equation}
\begin{bmatrix} z_0 \\ y_0 \\ x_0\end{bmatrix} \to \begin{bmatrix} \alpha+ (z_0-\alpha)(1-\lambda) \\ y_0\sqrt{1-\lambda} \\ x_0\sqrt{1-\lambda} \end{bmatrix}
\end{equation}
where $\alpha=(2p-1)$ is the $z$ coordinate of the fixed point of the thermalising map, {\it i.e.} of the Gibbs state at the same temperature as the reservoir, and it is linked to the decay constant $\eta$ as: $\eta=1/\alpha$.

It is commonplace to treat thermalisation in a rotating frame, making use of the interaction picture to dispense with the Hamiltonian evolution; this is also possible whenever the former is much slower than dissipation, $\omega\gg1/\eta$, realising an effective decoupling between unitary and dissipative dynamics.

{\it Evolution of the matrix elements.} The evolution of a density matrix $\rho$ under contractive dynamics is conveniently cast in terms of the evolution of its individual elements elements, distinguishing between populations and coherences. We assume that the contractive map can be described with a single parameter $\lambda \in [0,1]$, linked to the time $t$. In what follows, we consider the case of i) single systems, and ii) coupled systems. 

Case i) The density matrix of a single qubit, at a generic time parametrised as $\lambda$, is written as:
\begin{equation}
\rho(\lambda)=\psi(\lambda)+\rho_{\infty},
\end{equation} 
where $\rho_{\infty}$ represents the fixed point of the evolution. The varying part can be decomposed as:
\begin{equation}
\psi(\lambda)=\frac{1}{2}[\psi_p(0)(1-\lambda) +\psi_c(0) \sqrt{1-\lambda} ], 
\end{equation} 
where we have distinguished the initial populations $\psi_p(0)$, and the initial coherences $\psi_c(0)$. These obey the equations
\begin{equation}
\begin{aligned}
\psi_p(\lambda)=\psi_p(0)(1-\lambda) \ \  \ \  \ \   \ \   \ \  \dfrac{d\psi_p(\lambda)}{d\lambda}=-\frac{\psi_p(\lambda)}{(1-\lambda)}\\
\psi_c(\lambda)=\psi_c(0)\sqrt{1-\lambda} \ \  \ \  \ \   \ \   \ \  \dfrac{d\psi_c(\lambda)}{d\lambda}=-\frac{\psi_c(\lambda)}{2(1-\lambda)}
\end{aligned}
\end{equation} 
showing how different decay constants affect the behaviour of the population and the coherence elements, with the latter dominating the evolution at long times.

Case ii) With two partitions available, we can consider the evolution of the total state, including both subsystems. We can distinguish two cases: in the first, only the subsystem $A$ undergoes the contractive dynamics, while the other, $B$, is kept fixed: the applied map is then in the form $ {\lbrace \Phi_\lambda^p \rbrace}_A \otimes \mathbb{I}_B$. The matrix can be divided in blocks, each undergoing an evolution closely resembling the one described previously. It is worth remarking that  all terms on the diagonal of each block evolve with a decay constant $1-\lambda$, regardless them being coherences or populations. For these terms the equation of motion holds:
\begin{equation}
\psi_d(\lambda)=\psi_d(1)+(\psi_d(0)-\psi_d(1))(1-\lambda)
\label{eq:d}
\end{equation}
where $d=p$ for the genuine population terms, and $d=m$ for the terms in the off-diagonal blocks, which we call mixed.
Differently from the case of a single system, these diagonals need not satisfying a normalisation condition: the fixed point of the evolution is not unique, and will depend explicitly on the initial point. Due to the linearity of the equation of motion, the trajectories can not intersect: the motion will takes place on one specific surface out of a set of non-intersecting planes, each containing a fixed point.

The second case has both subsystems acted upon with a contractive map: the overall action can be written as the composition of two individual maps ${\lbrace \Phi_\lambda^p \rbrace}_A\otimes {\lbrace \Phi_\lambda^p \rbrace}_B$, both with a fixed point. This gives the evolution of the matrix elements in the form:
\begin{equation}
\begin{aligned}
\psi_p(\lambda)&=\psi_p(1)+[\psi_p(0)-\psi_p(1)-\psi_i](1-\lambda)+\psi_i(1-\lambda)^2,\\
\psi_m(\lambda)&=[\psi_d(1)+[\psi_m(0)-\psi_m(1)](1-\lambda)]\sqrt{1-\lambda},\\
\psi_c(\lambda)&=\psi_c(1)(1-\lambda),
\end{aligned}
\end{equation}
where the diagonal terms in \eqref{eq:d} now undergo different evolutions depending if they are populations $\psi_p(\lambda)$ or coherences $\psi_m(\lambda)$, and $\psi_i$ is a term depending on the local populations as well as on the fixed points. These expressions show how the speed of the coherences is increased, while that of the populations remain approximately the same in the limit of long times $\lambda\to 1$. In general, for $N$ coupled qubits the equation of motions for the mixed terms are in the form
\begin{equation}
\psi_{j}^n(\lambda)=\sum_{k=1}^n[h_{j,k}(1-\lambda)^k](1-\lambda)^{(N-n) \over 2}
\label{eq:y_sist_generico}
\end{equation} 
where $n$ counts the number of subsystems for which $\psi_{m}^n(\lambda)$ behaves as a population, Eq.\eqref{eq:d}. Its time derivative is expressed as  
 \begin{equation}
\dfrac{d\psi_j^n(\lambda)}{d\lambda}=\frac{f(\{\psi_l^k(\lambda)\}, \alpha, \lambda
)}{(1-\lambda)} 
\label{eq:dx_(1-lambda, lambda)}
\end{equation} 
where $\alpha$ are the coordinates of the fixed point, and $f$ is a limited function, which, in general, is not identically zero. For the Markovian case, we necessarily have:
 \begin{equation}
\dfrac{d\psi_j^n(\lambda)}{d\lambda}=\frac{f(\{\psi_l^k(\lambda)\}, \alpha
)}{(1-\lambda)}, 
\label{eq:dx_(1-lambda)}
\end{equation} 
with no explicit dependence on the decay parameter $\lambda$.

{\it General considerations on the evolution of the metric.} We start with some generic remarks on the statistical speed $V(\rho)^2=G_{t t}=\sum_{ij} G_{\chi_i\chi_j}\dfrac{d\chi_i(t)}{dt}\dfrac{d\chi_j(t)}{dt}$. Unless the evolution depends explicitly on time, a speed can be associated univocally to a state at a given time: this becomes a representative parameter of the state itself. Further, as a consequence of \eqref{eq:dx_(1-lambda, lambda)}, we generally have $G_{\lambda \lambda} =  g_{\lambda \lambda}(1-\lambda)^{-2} \rightarrow \infty$ for $\lambda \to1$, except for specific initial conditions; in the Markovian limit \eqref{eq:dx_(1-lambda)}, $g_{\lambda \lambda}$ is purely geometric. Such a divergence is inherent to looking at the evolution in terms of the decay parameter, rather than of proper time; in fact, this is removed, for instance, for an exponential decay. If $\lambda=1-e^{-t \eta}$ then we have $G_{tt}=\eta^2 g_{\lambda \lambda} $. For these reasons, $g_{\lambda \lambda}$ is a preferable quantity to analyse.

It can be demonstrated that:
 \begin{itemize}
\item the field of quadratic speeds is convex for stochastic processes. This descends from the convexity of the distances that generate the metric by means of infinitesimal displacements;
\item speeds are monotonically decreasing for Markov processes. This is ensured by the fact that $\Gamma_{T(\rho_0)}\left[T(d_t\rho_0),T(d_t\rho_0)\right]=\Gamma_{T(\rho_0)}\left[d_tT(\rho_0),d_tT(\rho_0)\right]=V(\rho)^2$; while this does not hold in general, it can be verified to hold true under the application of the Chapman-Kolgomorov transformation.
 \end{itemize}
 Since a unitary transformation $U$ and its inverse $U^{-1}$ are examples of Markovian maps, the following invariance condition holds:
\begin{equation}
\label{G_Invarinte_U}
\Gamma_{\rho}(d\rho,d\rho)=\Gamma_{U\rho U^{\dagger}}(Ud\rho U^{\dagger}, Ud\rho U^{\dagger})
\end{equation}
Furthermore, if $U$ is diagonal in the energy basis, under thermalisation we have $Ud\rho U^{\dagger} = d \left(U \rho U^{\dagger} \right)$, thus:
\begin{equation}
G_{\lambda \lambda}(\rho)=G_{\lambda \lambda}(U \rho U^{\dagger}).
\end{equation}

This relation has the important consequence that the speed with respect to the parameter $\lambda$ is the same in both Schr\"odinger and interaction pictures. However, in the former case, one needs to consider the contribution coming from the Hamiltonian part: in this case, the total speed can be found from the relation 
\begin{equation}
V(\rho)^2 dt^2=\Gamma_{\rho}(d\rho_{\lambda},d\rho_{\lambda})+\Gamma_{\rho}(d\rho_{\lambda},d\rho_{t})+\Gamma_{\rho}(d\rho_{t},d\rho_{t})
\end{equation}
considering at once the increments $d\rho_\lambda$ from dissipation and $d\rho_t$ coming explicitly from the pure Hamiltonian. 
Remarkably, for single qubits the two evolutions are orthogonal, and the speed can be decomposed as $V^2=V^2_\lambda + V^2_t$, since the rotational elements of $d\rho_{\lambda}$ e $d\rho_{t}$ are orthogonal in the Euclidean space. However, this is not the case in higher dimensions, as we will see.

{\it Metric for the single qubit.} In any evolution occurring on a curve parametrised by the set $\{ \lambda_i \}$, the density matrix is identified by its coordinate $\{x(\lambda_i), y(\lambda_i), z(\lambda_i)\}$, thus $G_{\lambda_i,\lambda_j}=\sum_{l,k=x,y,z}G_{l,k}\frac{dl}{d\lambda_i}\frac{dk}{d\lambda_j}$. Given the invariance of the Kraus operators of the GAD channel under rotations around $z$, the speed is better expressed in polar coordinates $\{r, \theta, \phi \}$, and their increments in the radial ($dr$) and transverse ($dn$) directions; based on the definition \eqref{ds_contrattive}, we obtain
\begin{equation}
\label{G_Sqbit-polari}
\begin{split}
\Gamma_\rho
=\frac{1}{4}\bigg{(}\frac{dr^2}{1-r^2}
+\frac{c^{f}(r)}{2}dn^2\bigg{)}
\end{split}
\end{equation}
where $c^{f}(r)$ is a Morozova-\v Cencov function, comprised between the minimal function $c^{m}(r)=2$, and the maximal function $c^{M}(r)=\frac{2}{1-r^2}$.

The {\it maximal} metric, in the sense that it generates the maximal distance, corresponds to $c^M(r)$ and is written as 
\begin{equation}
\label{G_Sqbit_M}
\begin{split}
\Gamma_{\rho}^M
=\frac{1}{4}\frac{dz^2+dx^2+dy^2}{(1-r^2)}
\end{split}
\end{equation}
which is the standard Euclidean increment, weighted by a constant curvature; this is peculiar to qubits, and does not extend to general dimensions. $\Gamma_\rho^F$ is the minimal metric and is associated to the quantum Fisher information:
\begin{equation}
\label{G_Sqbit_fish}
\begin{split}
\Gamma_\rho^F
&=\frac{1}{4}\left( dz^2+dx^2+dy^2+\frac{d(r^2)^2}{(1-r^2)}\right). 
\end{split}
\end{equation}

\begin{figure}
\includegraphics[width=0.84\columnwidth]{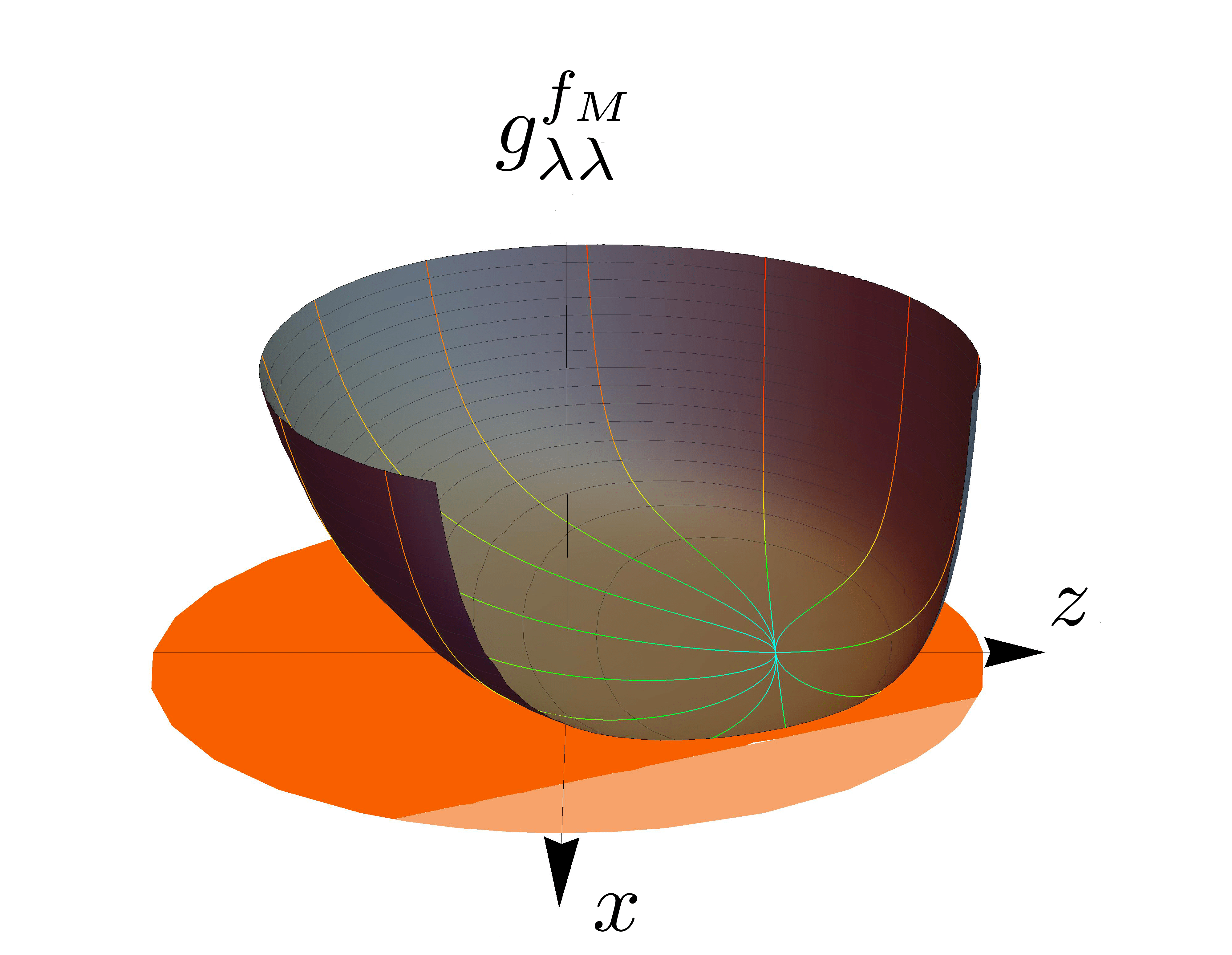}
\includegraphics[width=0.14\columnwidth]{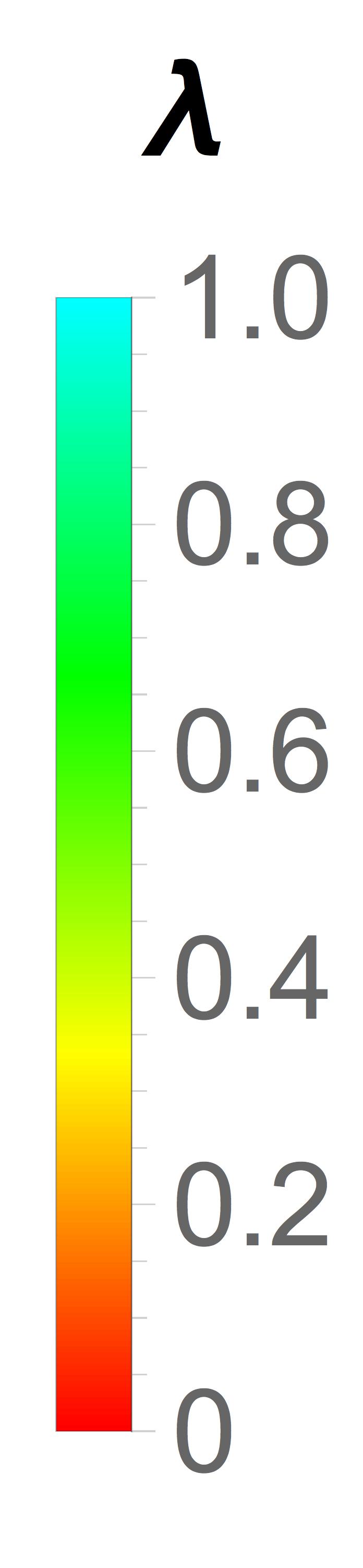}
\caption{3D representation of the speed $G_{tt}$ as a function of the $x$ and $z$ component of the initial state. Speeds along the trajectories are also marked, with the parameter $\lambda$ indicated according to the colour code in the legend.}
\label{daddy}
\end{figure}

We now turn our attention to the speed, referring only to that associated to dissipation; the Hamiltonian contribution is simply $V_{t}(\rho)^2=\frac{\omega^2}{4}(x^2+y^2)$.
For the thermalisation, we can write
\begin{equation}
\label{G_Sqbit_GAD_fish}
\begin{split}
&g^F_{\lambda\lambda}(\vec{r}(\lambda),\alpha)=\\
&=\frac{1}{4}\frac{(z(\lambda)-\alpha)^2+(x(\lambda)/2)^2[1-(z(\lambda)-2\alpha)^2]}{1-r(\lambda)^2},
\end{split}
\end{equation}
where we have set the reference frame so that $y(0)=0$. Its qualitative behaviour is represented in Fig.\ref{daddy}: points further from the fixed point move with different speeds, however an asymmetry between the ``population" and ``coherence" components is present.

{\it Metric for two-qubit systems.} The phenomenology of the single qubit is sufficiently clear, mostly thanks to its low dimensionality. However, this carries peculiarities, such as the fact that the coordinate representation defines a sphere, or that the natural parametrisation relies on a set of mutually anticommuting matrices; none of these are maintained in higher dimensions. We then consider what results from the analysis of two-qubit states; this adds an extra ingredient with respect the previous case, in that quantum correlations can be present. While the problem could be explored by means of exact expressions, their explicit form would be hard to interpret due to the presence of large number of parameters. Our strategy to obtain an insight is to discuss characteristic examples, and adopt numerics.

A 4$\times$4 density matrix is written as
 \begin{equation} 
\label{matx4x4}
\begin{split}
\rho=\frac{1}{4}
\begin{bmatrix}
1+z_0 &  y_ {b0} & y_ {c0} &  x_0 \\
y_{b0}^*&  1+z_1  & x_1^* &  y_ {c1} \\
y_ {c0}^* &  x_1 & 1+z_2 &  y_ {b1} \\
x_0 ^*&  y_ {c1}^*  & y_ {b1}^* &  1+z_3
\end{bmatrix},
\end{split}
\end{equation}
where the terms $z_i$ describe the populations, the terms $y_{i}$ the local coherences, and the terms $x_i$ the coherence arising from correlations between the two subsystems. For the sake of simplicity, we will investigate  the following cases:

\begin{equation} 
\label{phi_deg}
\begin{split}
\rho_{Yb}=\frac{1}{4}
\begin{bmatrix}
1+z_0 &  y_{b0}  & 0 &  0 \\
y_{b0}^* &  1+z_1  & 0 &  0 \\
0 &  0  & 1+z_2 &  y_{b1} \\
0 &  0  & y_{b1}^* &  1+z_3 
\end{bmatrix} \\
\rho_{Yc}=\frac{1}{4}
\begin{bmatrix}
1+z_0 &  0  & y_{c0} &  0 \\
0 &  1+z_1  & 0 &  y_{c1} \\
y_{c0}^* &  0  & 1+z_2 &  0 \\
0 &  y_{c1}^*  & 0 &  1+z_3 
\end{bmatrix} \\
\rho_{X}=\frac{1}{4}
\begin{bmatrix}
1+z_0 &  0  & 0 &  x_0 \\
0 &  1+z_1  & x_1^* &  0 \\
0 &  x_1  & 1+z_2 &  0 \\
x_0^* &  0  & 0 &  1+z_3 
\end{bmatrix} \\
\end{split}
\end{equation} 
These have been chosen by reason of their being divided in blocks: this simplifies their geometry, while covering a broad class of phenomena. Indeed, they present coupled dynamics between populations and different kinds of coherence, and their evolution maps among states of the same type. We remark how entanglement can only be present in the $X$ class.

We can recognise a universal behaviour in the evolution of the three classes, by adopting the definitions:
\begin{equation}
\label{parametri_XYZ}
\begin{aligned}
z_a=z_{01},\, z_b= z_{23}, \, x_a=y_{b0}, \, x_b= y_{b1},\, \Delta=\delta z_{01},\\
z_a=z_{02}, \, z_b= z_{13}, \, x_a=y_{c0}, \, x_b= y_{c1},\, \Delta=\delta z_{02},\\
z_a= z_{03}, \,  z_b= z_{21}, \, x_a=x_{0}, \, x_b= x_{1}^*,\, \Delta=\delta z_{03}, 
\end{aligned}
\end{equation}
respectively, for the states in \eqref{phi_deg}; we have used the shorthand notation $\chi_{ij}=(\chi_i-\chi_j)/2$, $\delta\chi_{ij}=(\chi_i+\chi_j)/2$. These isolate two geometric partitions - which do not correspond to the constituent qubits; this is also shown by the eigenvalues of the density matrix, that reads 
\begin{equation}
\begin{aligned}
p_{a,1}=\frac{1}{4}( \delta_a + \xi_a),\,p_{a,2}=\frac{1}{4}( \delta_a - \xi_a),\\
p_{b,1}=\frac{1}{4}( \delta_b + \xi_b),\,p_{b,2}=\frac{1}{4}( \delta_b - \xi_b),
\end{aligned}
\end{equation}
where $\delta_a=1+\Delta$ and $\delta_b=1-\Delta$ are the probabilities of being in either partition, and $\xi_u=z_u^2+x_u^2$ $(u=a,b)$. While a direct interpretation is not straightforward in the general case, 
for $Y_b$ states $\{z_a,x_a\}$ and $\{z_b,x_b\}$ are population and coherences of the first qubit, conditioned on a measurement of the second in its ground or excited state, respectively. The converse is true for $Y_c$ states.

For such matrices, the metric can be decomposed in two parts: despite the symmetry in the geometric aspects, the actual dynamics differs in the three cases, due to the time dependence discussed in the previous sections. The behaviour of each partition resembles that of a single qubit, despite the fact that $a$ and $b$ do not refer to physical subsystems:
	\begin{equation}
\begin{aligned}
G_{\lambda,\lambda}&=G^a_{\lambda,\lambda}+G^b_{\lambda,\lambda},\\
G^u_{\lambda,\lambda}&=F^u_{\lambda,\lambda}+Q^u_{\lambda,\lambda},\\
F^u_{\lambda,\lambda}&={1 \over 8}{\delta_u \left[  \dot{\delta}_u^2+ \dot{\xi}_u^2 \right] -2 \xi_u \left[ \dot{\delta}_u  \dot{\xi}_u \right]  \over \delta_u^2 -\xi_u^2 },\\
Q^u_{\lambda,\lambda}&={c^f(p_{u,1},p_{u,2}) \over 32 \xi_u^2}\left(x_u \dot{z}_u - z_u \dot{x}_u  \right).
\end{aligned}
\end{equation}
The compact notation $\dot \chi=\partial_\lambda \chi$ is used. In the maximal metric, the resulting expressions are:
\begin{equation}
\label{G_Dqbit_bloc_m}
\begin{aligned}
&G^{u,M}_{\lambda\lambda}
=\frac{\delta_u}{8}\frac{\dot{z_u}^2+\dot{x_u}^2+\dot{\delta}_u^2}{( \delta_u^2 - \xi_u^2)}
-\frac{1}{4}\frac{\dot{\delta_u}\left(\dot{z_u}z_u+\dot{x_u}x_u\right)}{( \delta_u^2 - \xi_u^2)},
\end{aligned}
\end{equation}
These can be rearranged in order to isolate a term depending on $\dot{\Delta}$, {\it i.e.} the change in the probability of being in either partition:
\begin{equation}
\label{G_Dqbit_bloc_m1}
\begin{aligned}
&G^{M}_{\lambda\lambda}
=\frac{\dot{\Delta}}{8}D^{M}_{\lambda\lambda}\left(\delta_a^2 - \xi_a^2, \delta_b^2 - \xi_b^2\right)+\sum_{i=a,b} \frac{\delta_i}{8} \frac{\dot{z_i}^2+\dot{x_i}^2}{\delta_i^2-\xi_i^2} \\
&D^{M}_{\lambda\lambda}(X,Y)=\dot{\log}{\Big{[}\frac{X}{Y}\Big{]}}- 2 \dot{\Delta} \left(\frac{1}{X}+\frac{1}{Y}\right) - 2 \dot{\Delta}\Delta \left(\frac{1}{X}-\frac{1}{Y}\right).
\end{aligned}
\end{equation}
The first two terms resemble an average speed on the partitions for fixed $\delta_a$ and $\delta_b$. The third term is the one linked to the transfer probability between the two partions $a$ to $b$, and, in general, it can always be rearranged in such a way to depend only on the determinants of the blocks $D_u=\delta_u^2-\xi_u^2$. In the Fisher metric we can write: 

\begin{equation}
\label{G_Dqbit_blocchi_fish}
G^{F}_{\lambda\lambda}
=\sum_{i=a,b} \frac{\left( \dot{z_i}^2+\dot{x_i}^2 \right)}{8 \delta_i} + \frac{1}{32 \delta_i} \frac{\dot{D_i^2}}{D_i}. 
\end{equation}

In what follows, we refer to the single qubit property of the quadratic speed in the Schrodinger scheme to be severable in two contributions namely the quadratic speed in the interaction scheme, and the one associated with the Hamiltonian H. When moving in the two-qubit scenario, such      a peculiarity does not immediately translate in an universal relation as it is restricted to some {\it special} cases. We summarize the range of applicability of the speed decomposition in the subsequent list: i) for the infinitesimal transformation $\mathbb{I}_A \otimes U_B {\lbrace \Phi_\lambda^p \rbrace}_B$, the speed for the $X$ states is severable (Y), and the same holds for the states $Y_b$ (Y) and $Y_c$ (Y); ii) for the infinitesimal transformation $U_A \mathbb{I}_A \otimes {\lbrace \Phi_\lambda^p \rbrace}_B$: $X$ (Y), $Y_b$ (Y), $Y_c$ (N); iii) for the infinitesimal transformation ${\lbrace \Phi_\lambda^p \rbrace}_A \otimes U_B {\lbrace \Phi_\lambda^p \rbrace}_B$: $X$ (Y), $Y_b$ (N), $Y_c$ (Y); iv) for the infinitesimal transformation $U_A {\lbrace \Phi_\lambda^p \rbrace}_A \otimes U_B$: $X$ (Y), $Y_b$ (Y), $Y_c$ (N).

\end{document}